\documentclass[useAMS,usenatbib]{mn2e}

\usepackage{latexsym,graphicx,natbib}

\newcommand\kms{{\rm\,km\,s^{-1}}}
\newcommand\msun{\rm\,M_\odot}

\def\apgt{\ {\raise-.5ex\hbox{$\buildrel>\over\sim$}}\ }
\def\aplt{\ {\raise-.5ex\hbox{$\buildrel<\over\sim$}}\ }

\title[A new Wolf-Rayet star in Cygnus]{Discovery of a new Wolf-Rayet star and its ring nebula in Cygnus}
\author[V.V.Gvaramadze et al.]
       {V. V.~Gvaramadze,$^{1}$\thanks{E-mail: vgvaram@mx.iki.rssi.ru (VVG); fabrika, olga, azamat@sao.ru
       (SF, OS, AFV); wrh, lida@astro.physik.uni-potsdam.de (WRH, LMO); goray, cher@sai.msu.ru (VPG, AMC);
        bomans@astro.ruhr-uni-bochum.de (DJB)} S.~Fabrika,$^{2}$
        W.-R.~Hamann,$^{3}$ O.~Sholukhova,$^{2}$
        \newauthor
        A. F.~Valeev,$^{2}$ V. P.~Goranskij,$^{1}$ A. M.~Cherepashchuk,$^{1}$, D. J.~Bomans$^{4}$
        \newauthor
        and L. M.~Oskinova$^{3}$\\
        $^{1}$Sternberg Astronomical Institute, Moscow State University, Universitetskij Pr. 13, Moscow 119992, Russia\\
        $^{2}$Special Astrophysical Observatory, Nizhnij Arkhyz, 369167, Russia\\
        $^{3}$Institute for Physics and Astronomy, University Potsdam, 14476 Potsdam, Germany\\
        $^{4}$Astronomical Institute, Ruhr-University Bochum, Universit\"{a}tstr. 150, 44780 Bochum, Germany
       }
\begin{document}

\date{Accepted 2009 August 3, Received 2009 July 20; in original form 2009 June 3}


\maketitle

\label{firstpage}

\begin{abstract}
We report the serendipitous discovery of a ring nebula around a
candidate Wolf-Rayet (WR) star, HBHA\,4202-22, in Cygnus using the
{\it Spitzer Space Telescope} archival data. Our spectroscopic
follow-up observations confirmed the WR nature of this star (we
named it WR\,138a) and showed that it belongs to the WN8-9h subtype.
We thereby add a new example to the known sample of late WN stars
with circumstellar nebulae. We analyzed the spectrum of WR\,138a by
using the Potsdam Wolf-Rayet (PoWR) model atmospheres, obtaining a
stellar temperature of 40\,kK. The stellar wind composition is
dominated by helium with 20 per cent of hydrogen. The stellar
spectrum is highly reddened and absorbed ($E_{B-V} = 2.4$\,mag, $A_V
= 7.4$\,mag). Adopting a stellar luminosity of $\log L/L_{\odot} =
5.3$, the star has a mass-loss rate of $10^{-4.7} \, \msun \, {\rm
yr}^{-1}$, and resides in a distance of 4.2 kpc. We measured the
proper motion for WR\,138a and found that it is a runaway star with
a peculiar velocity of $\simeq 50 \, \kms$. Implications of the
runaway nature of WR\,138a for constraining the mass of its
progenitor star and understanding the origin of its ring nebula are
discussed.
\end{abstract}


\section{Introduction}
\label{sec:intro}

It is believed that single stars of solar metallicity with an
initial mass of $\ga 20-25 \, \msun$ end their lives as Wolf-Rayet
(WR) stars (Vanbeveren, De Loore \& Van Rensbergen 1998; Meynet \&
Maeder 2003). The relatively small number of massive stars and the
short duration of the WR phase ($\la 0.5$ Myr) makes the WR stars
rare objects. The present total number of WR stars in the Galaxy
does not exceed several thousands (e.g. Shara et al. 1999; van der
Hucht 2001). The known population of WR stars is, however, an order
of magnitude less numerous (van der Hucht 2006). The large disparity
between the expected and the observed number of WR stars is mostly
caused by the huge interstellar extinction in the Galactic plane,
which strongly limits the traditional method for searching for WR
stars using optical surveys. In this respect, the infrared (IR)
observations are of high importance since the interstellar medium
(ISM) is much more transparent in the IR compared with the optical.

The effectiveness of IR observations in identifying new WR stars was
demonstrated by Shara et al. (2009) who carried out a near-IR
narrow-band imaging survey of the inner Galactic plane to select
candidate WR stars and discovered 41 new WR stars in spectroscopic
follow-ups (cf. Homeier et al. 2003; Hadfield et al. 2007; Mauerhan,
Van Dyk \& Morris 2009). Another possible way to search for new WR
(and other evolved massive) stars is through detection of their (IR)
circumstellar nebulae, created after a massive star leaves the main
sequence (e.g. Marston et al. 1999; Egan et al. 2002; Clark et al.
2003).

In this paper, we report the discovery of a new WN8-9h star in
Cygnus via the detection of its IR nebula. We thereby add a new
example to the known sample of late WN (WNL) stars with
circumstellar nebulae (e.g. Esteban et al. 1993). Inspired by this
discovery, we undertook an extensive search for similar objects
using the archival data of the Spitzer Legacy
Programs\footnote{http://irsa.ipac.caltech.edu/Missions/spitzer.html}
and discovered many dozens of ring-like and bipolar IR shells
(Gvaramadze, Kniazev \& Fabrika 2009; cf. Carey et al. 2009).
Spectroscopic follow-ups of central stars associated with two dozens
of the shells showed that they are either candidate Luminous Blue
Variables or related (WNL, blue supergiant) stars (Gvaramadze et al.
2009, in preparation; Kniazev et al. 2009, in preparation), which
confirmed that the IR imaging provides a powerful tool for revealing
evolved massive stars via detection of their circumstellar nebulae.

\section{A new ring nebula and its central star}
\label{sec:neb}

The nebula, which is the subject of this paper, was discovered
serendipitously during our search for bow shocks around OB stars
running away from star clusters in the Cygnus X region (for
motivation of this search see Gvaramadze \& Bomans 2008a). The
nebula was detected in the archival data of the {\it Spitzer Space
Telescope}, obtained with the Infrared Array Camera (IRAC) and the
Multiband Imaging Photometer for {\it Spitzer} (MIPS) (Werner et al.
2004) within the framework of the Cygnus-X Spitzer Legacy
Survey\footnote{http://www.cfa.harvard.edu/cygnusX/index.html}.

\begin{figure}
\begin{center}
\includegraphics[width=1.0\columnwidth,angle=0]{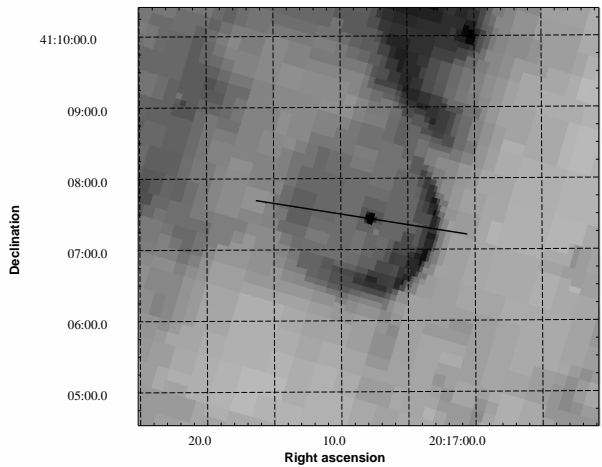}
\end{center}
\caption{
{\it Spitzer} MIPS $24 \, \mu$m image of a new ring nebula in Cygnus and its
    central star HBHA\,4202-22 (WR\,138a). The orientation of the spectrograph slit in our
    optical observation is indicated by a line.
}
\label{fig:neb}
\end{figure}

Fig.\,\ref{fig:neb} shows the MIPS $24 \, \mu$m image of the nebula
and its central star. The nebula has a clear ring-like structure
with a diameter of $\simeq 2.3$ arcmin and enhanced brightness along
the west rim. The central star is offset for $\sim 0.2$ arcmin from
the geometric centre of the nebula, being closer to the brightest
portion of the nebula (a possible origin of the nebula and its
brightness asymmetry are discussed in Section\,\ref{sec:nebula}).
The Cygnus-X Spitzer Legacy Survey also provides IRAC $3.6, 5.8 \,
\mu$m and MIPS $70 \, \mu$m images of the field containing the
nebula; none of them shows its signatures.

The optical counterpart to the central star was identified by
Dolidze (1971) as a possible WR star due to the presence in its
spectrum of an emission band around $\lambda 6750$ \AA. This star
was included in the Catalogue of H-alpha emission stars in the
northern Milky Way (Kohoutek \& Wehmeyer 1999) as a candidate WR
star (named in the SIMBAD database as [KW97]\,44-47 or
HBHA\,4202-22), but was not listed in the VIIth Catalogue of
Galactic Wolf-Rayet Stars (van der Hucht 2001) nor in the annex to
this catalogue (van der Hucht 2006). Recently, Corradi et al. (2008)
put this star in the list of candidate symbiotic stars on the basis
of its IPHAS and 2MASS colours.

\section{Observations}
\label{sec:obs}

\subsection{Spectroscopy}

To verify the nature of HBHA\,4202-22, we had performed
spectroscopic observations with the Russian 6-m telescope using the
SCORPIO\footnote{Spectral Camera with Optical Reducer for
Photometrical and Interferometrical Observations;
http://www.sao.ru/hq/lsfvo/devices/scorpio/scorpio.html} focal
reducer (Afanasiev \& Moiseev 2005) in a long-slit mode with a slit
width of 1 arcsec, providing a spectral resolution of 5 \AA. The
orientation of the (6 arcmin long) slit is shown in
Fig.\,\ref{fig:neb}. The spectra were taken on two occasions: 2008
October 10 (spectral range $\lambda\lambda 4000-5700$~\AA) and 2008
November 11 (spectral ranges $\lambda\lambda 4000-5700$~\AA \, and
$\lambda\lambda 5700-7500$~\AA), and reduced using standard
procedures.

\subsection{Photometry}

Using secondary photometric standards in the regions of symbiotic
variable stars V407 Cyg and AS 323 published by Henden \& Honeycutt
(1997), we determined accurate magnitudes of four comparison stars
near HBHA\,4202-22. The observations were carried out with the
Special Astrophysical Observatory (SAO) 1-m Zeiss reflector and
$UBVR_{\rm C}I_{\rm C}$ photometer in the good photometric night on
2009 May 29. The photometry accuracy is better than 0.02 mag. We
have made expanded CCD standard star set of 18 stars to measure a
DSS POSS-1 blue image. HBHA\,4202-22 was measured using POSS-O plate
taken in the epoch of 1954.57, CCD frames taken with the Russian 6-m
telescope in 2008.90 and CCD frames taken with SAO 1-m Zeiss
telescope in 2009. The results are given in Table\,\ref{tab:res}.

\begin{table*}
  \caption{Photometry of HBHA\,4202-22 (WR\,138a)}
  \label{tab:res}
  \begin{center}
  \begin{minipage}{\textwidth}
    \begin{tabular}{ccccc}
      \hline
      \hline
      JD & $B$ & $V$ & $R_{\rm C}$ & Source \\
      \hline
      2434951 & $17.15\pm0.06$ & -- & -- & POSS-1 \\
      2454796 & $17.25\pm0.01$ & $15.44\pm0.01$ & -- & 6-m/SCORPIO \\
      2454981 & $17.34\pm0.02$ & $15.57\pm0.02$ & $13.90\pm0.02$ & 1-m Zeiss \\
      2455008 & $17.27\pm0.02$ & $15.53\pm0.02$ & $13.85\pm0.02$ & 1-m Zeiss \\
            \hline
    \end{tabular}
    \end{minipage}
    \end{center}
\end{table*}

The uniform 1-m Zeiss photometry shows that the object is variable
with amplitudes between 0.04 and 0.07 in different filters in the
time scale of a month. A total variability range in B filter between
1955 and 2009 is 0.19 mag. This level of variability is
typical of WNL stars (e.g. 
Marchenko et al. 1998).

To measure the {\it Spizer}'s IRAC and MIPS fluxes from
HBHA\,4202-22, we performed aperture photometry of the star using
the {\tt MOPEX/APEX} source extraction package and applied aperture
correction, obtained from the nearby bright point sources. The
fluxes are listed in Table\,\ref{tab:obs}; the estimated errors are
$\sim 2-3$ per cent. For the flux at $70 \, \mu$m, we give a
$3\sigma$ upper limit since HBHA\,4202-22 was not detected at this
wavelength.

The details of HBHA\,4202-22 are summarized in Table\,\ref{tab:obs}.
The coordinates and the $J,H,K_s$ magnitudes are taken from the
2MASS All-Sky Catalog of Point Sources (Skrutskie et al. 2006).
\begin{table}
  \caption{Details of HBHA\,4202-22 (WR\,138a)}
  \label{tab:obs}
  \begin{center}
  \begin{minipage}{\textwidth}
    \begin{tabular}{ccccc}
      \hline
      \hline
      RA(J2000) &  $20^{\rm h} 17^{\rm m} 08\fs12$ \\
      Dec.(J2000) &  $41\degr 07\arcmin 27\farcs0$ \\
      $l,b$ & 78.3203, 3.1536 \\
      $B$ (mag) &  $17.25\pm0.01$ \\
      $V$ (mag) &  $15.44\pm0.01$ \\
      $J$ (mag) & $10.15\pm0.02$ \\
      $H$ (mag) & $9.27\pm0.02$ \\
      $K_s$ (mag) & $8.65\pm0.02$ \\
      $[3.6]$ (mJy) & $168.0\pm5.0$ \\
      $[5.8]$ (mJy) & $154.0\pm4.6$ \\
      $[24]$ (mJy) & $45.7\pm1.4$ \\
      $[70]$ (mJy) & $< 28.1$ \\
      \hline
    \end{tabular}
    \end{minipage}
    \end{center}
\end{table}

\subsection{Proper motion}
\label{sec:obs-prop}

We measured the proper motion for HBHA\,4202-22 using four epochs in
the POSS-1 and POSS-2 Digitized Sky Surveys and the epoch of our
6-m/SCORPIO frames. The full epoch range is 54.3 years. 21 reference
stars located within an angular radius of 2.5 arcmin from the object
position were used. Typical rms deviations of reference stars in
these frames are between 0.21 and 0.56 arcsec. Linear trends are
well seen in both coordinates. To evaluate the stellar proper motion
and its errors, we got a linear approximation of these trends using
the method of least squires and derived mean rms deviations of
stellar coordinates from these linear approximations. The proper
motion errors were estimated by dividing these mean deviations by
the epoch difference. The result is $\mu _\alpha = -4.9\pm 1.2 \,
{\rm mas} \, {\rm yr}^{-1}$, $\mu _\delta =-4.7\pm 1.2 \, {\rm mas}
\, {\rm yr}^{-1}$.

To convert the observed proper motion into the true tangential
velocity of the star, we use the Galactic constants $R_0 =8$ kpc and
$\Theta =200 \, \kms$ (e.g. Reid 1993; Avedisova 2005) and the solar
peculiar motion ($U_{\odot} , V_{\odot} , W_{\odot}) = (10.00, 5.25,
7.17) \, \kms$ (Dehnen \& Binney 1998), and adopt a distance $d=4.2$
kpc (see Section\,\ref{sec:dist}). We found that the star is moving
in the west-southwest direction with a peculiar (transverse)
velocity of $\simeq 50\pm 24 \kms$ (here we assume that the errors
are mainly due to the errors in our proper motion measurements).
Note that the vector of stellar peculiar velocity is oriented by
chance almost exactly along the spectrograph slit (see
Fig.\,\ref{fig:neb}).

\section{A new Wolf-Rayet star -- WR\,138a}

\subsection{Spectral type}
\label{sec:type}

\begin{figure}
\includegraphics[width=0.7\columnwidth,angle=270]{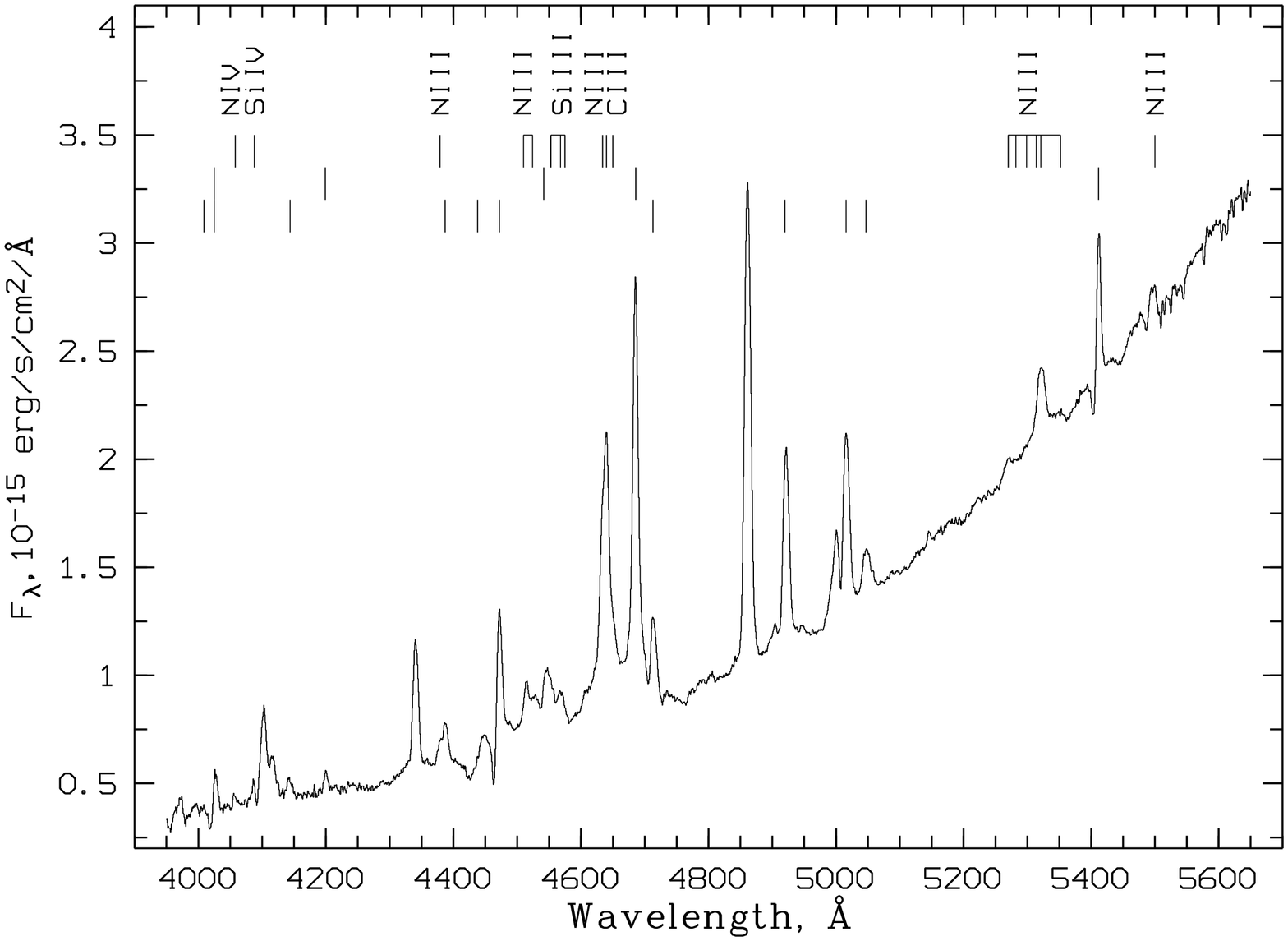}
\includegraphics[width=0.7\columnwidth,angle=270]{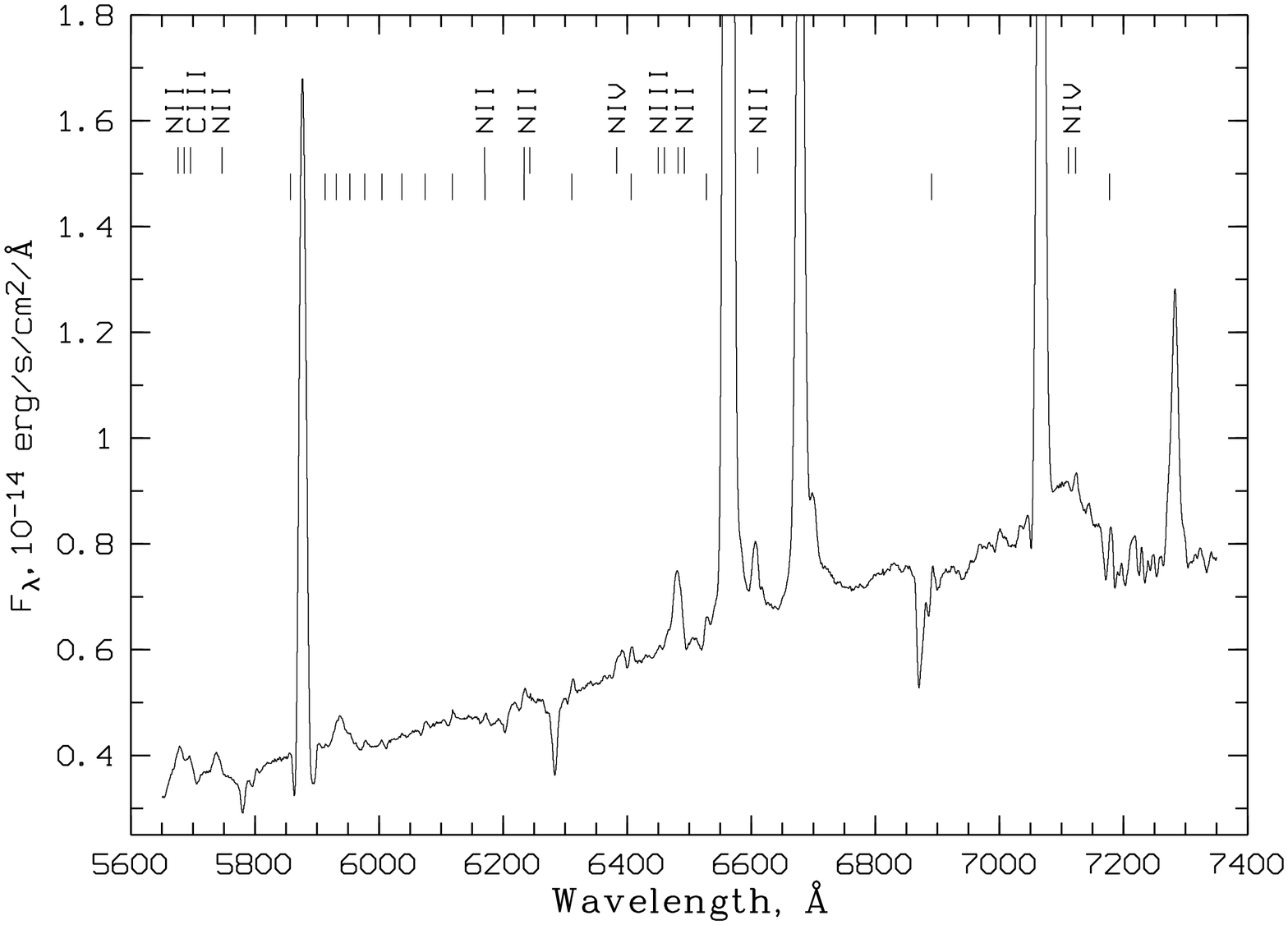}
\caption{Blue (top) and red (bottom) spectra of HBHA\,4202-22 (WR\,138a)
with some lines indicated (upper row). In the blue spectrum, the bottom row of
vertical bars marks numerous He\,{\sc ii} lines while the middle one marks
He\,{\sc i} lines. In the red spectrum, the vertical bars mark He\,{\sc ii}
lines, whereas very strong He\,{\sc i} lines are not marked.
H\,{\sc i} emissions are not marked on both spectra as well.
CCD fringes are notable at wavelengths longer than $\sim 6900$\,\AA.}
\label{fig:spec}
\end{figure}

In Fig.\,\ref{fig:spec} we present the spectrum of HBHA\,4202-22 in
the blue and red regions. The spectrum shows strong emission lines
of hydrogen, He\,{\sc i} and He\,{\sc ii}. We detect emissions of
N\,{\sc ii}, N\,{\sc iii}, C\,{\sc iii}, N\,{\sc iv}, Si\,{\sc iv}.
No forbidden lines can be seen in the spectrum. Many of the weaker
lines in the spectrum show P\,Cygni profiles, while the strongest
lines are entirely in emission. The strong absorptions visible in
the red spectrum are telluric. Numerous diffuse interstellar bands
(DIBs) are observed, in the blue region they are $\lambda\lambda
4428, 4726, 4762$, in the red the strongest one is at 6280 \AA.

Our spectra also show nebular emission lines along the whole length
of the slit. Radial velocity in H$\alpha$ line is constant along the
slit and equal to $-21\pm5 \, \kms$. From the Balmer decrement, we
estimated the interstellar extinction and obtained $A_V =
1.9\pm0.1$\,mag (also constant along the slit), which is much less
than $A_V = 7.4$\,mag derived from the stellar spectrum (see Section
4.4). Hence, we conclude that the nebular line emission originates
from the foreground and is not related to HBHA\,4202-22 and its ring
nebula.

The dominance of helium and nitrogen emission lines indicates that
HBHA\,4202-22 belongs to the nitrogen (WN) sequence of WR stars. The
presence of hydrogen emission lines is typical for WNL subtypes
(Hamann et al. 1991). We named this star WR\,138a, in accordance
with the numbering system of the VIIth Catalogue of Galactic
Wolf-Rayet Stars by van der Hucht (2001). To determine the subtype
of WR\,138a more precisely, we use the classification scheme
proposed by Smith (1968) and updated for WNL stars by Crowther,
Hillier \& Smith (1995) and Smith, Shara \& Moffat (1996). The
relative strengths of N\,{\sc iii} $\lambda \, 4640$, N\,{\sc iv}
$\lambda \, 4058$ and He\,{\sc ii} $\lambda \, 4686$ lines and the
P\,Cygni profiles of He\,{\sc i} lines suggest that the star belongs
to the WN8-9 subtype. The WN8-9 classification for WR\,138a also
follows from the position of this star on the diagrams showing a
comparison of the emission equivalent widths (EW) of He\,{\sc i}
$\lambda 5876$ versus He\,{\sc ii} $\lambda 4686$ lines and He\,{\sc
ii} $\lambda 4686$ EW versus FWHM for WNL stars in the Milky Way and
the Large Magellanic Cloud (LMC) (see Fig.\,1 of Crowther \& Smith
1997). With EW(5876)=40.0 \AA, EW(4686)=19.5 \AA \, and
FWHM(4686)=7.7 \AA, WR\,138a lies almost exactly on the line
separating the WN8 and WN9 stars. The further evidence for this
classification is given in Section\,\ref{sec:PoWR}. Following the
three-dimensional classification for WN stars by Smith et al.
(1996), we add a `h' suffix to WN8-9 to indicate the presence of
hydrogen emission lines, so that WR\,138a is a WN8-9h star.

\subsection{Radial velocity}
\label{sec:rad}

We have measured the radial velocities of the main emission lines in
the spectrum and the line widths corrected for the spectral
resolution. We used the Gaussian analysis for the measurements. The
accuracy of radial velocity measurements is $\la 10 \, \kms$, while
the relative velocities are measured with a notably better
precision.

For hydrogen lines H$\gamma$, H$\beta$ and H$\alpha$, we measured
heliocentric radial velocities $v_{\rm r} \simeq -10 \, \kms$ (FWHM
$\simeq 610 \pm 30 \, \kms$). The strongest He\,{\sc ii} $\lambda
4686$ line shows $v_{\rm r} \simeq -37 \, \kms$ (FWHM $\simeq 490 \,
\kms$) and it does not have the P\,Cyg profile. All He\,{\sc ii}
lines with the P\,Cyg profile have $v_{\rm r} \simeq 49 \, \kms$
(FWHM $\simeq 170 \, \kms$) and the radial velocity measured for the
absorption peak $v_{\rm a} \simeq -380 \, \kms$. Blue shifted
absorption obviously shifts emission to the red side and makes it
narrower. He\,{\sc i} lines show the same behaviour as He\,{\sc ii}
lines. The mean radial velocity for all `non-P\,Cyg' He\,{\sc i}
lines is $\simeq -11 \, \kms$ (FWHM $\simeq 590 \, \kms$).

We may adopt for the star's radial velocity $v_{\rm r} \sim -20 \,
\kms$, which is an average over the hydrogen lines, He\,{\sc ii}
$\lambda 4686$ and the `non-P\,Cyg' He\,{\sc i} lines. It is
necessary to note, however, that the radial velocity of the star
cannot be determined with good accuracy, because all the spectral
lines are formed in the wind. Resulting from a complicate formation
process in the expanding atmosphere, the emergent line profiles are
not exactly symmetric. Due to the high wind speed ($700 \, \kms$, as
found in our spectral analysis, see below), these effects are large
compared to possible radial velocity uncertainties.

\subsection{Spectral analysis and stellar parameters}
\label{sec:PoWR}

\begin{figure*}
\begin{center}
\includegraphics[width=\textwidth]{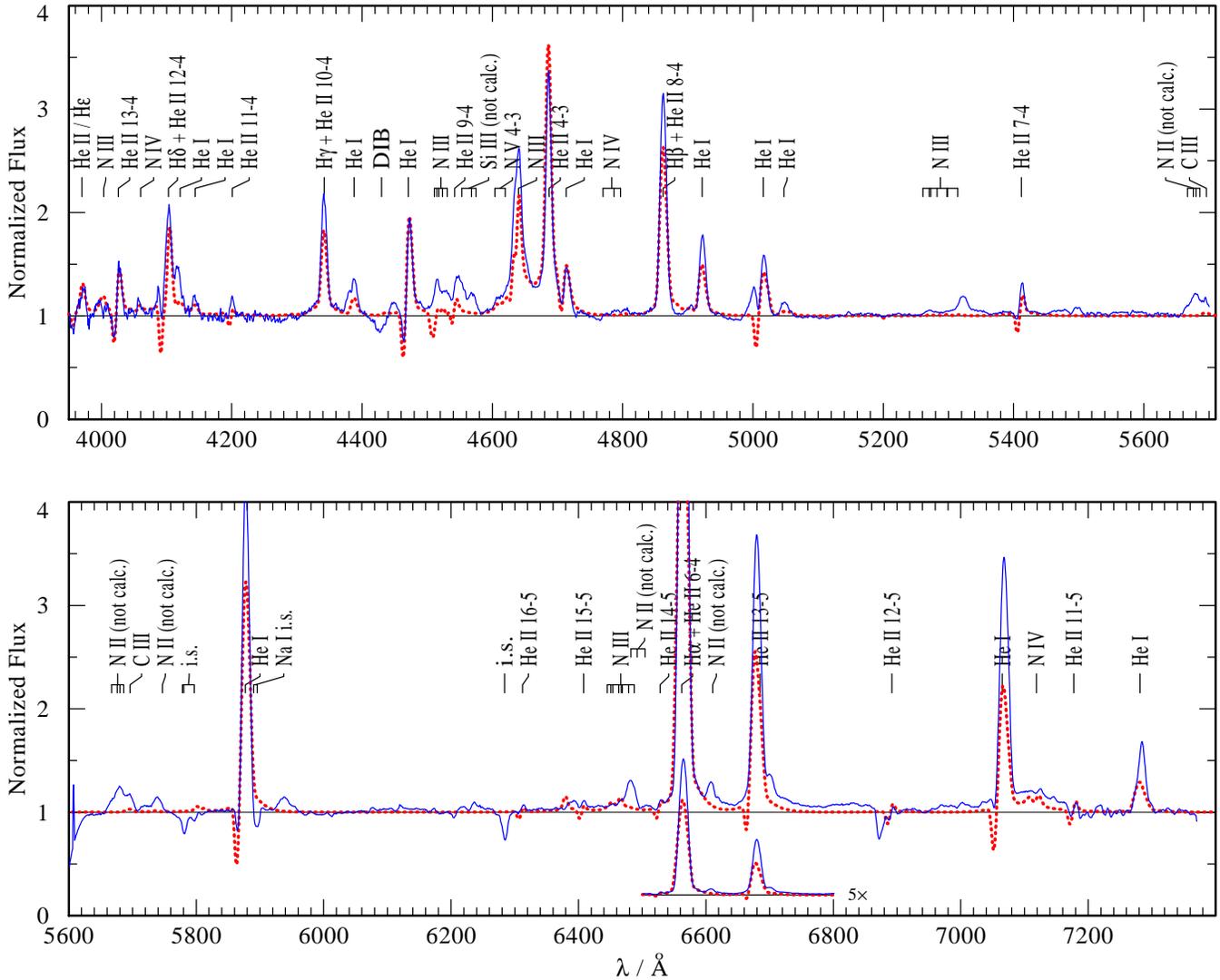}
\end{center}
\caption{Observed optical spectrum (blue/solid line) of WR\,138a,
compared with the best fitting model (red/dotted line) with the
parameters as given in Table\,\ref{tab:model}. The absolutely
calibrated observation had been divided by the reddened model
continuum for normalization.} \label{fig:linefit}
\end{figure*}

To analyze the stellar spectrum and to derive the fundamental
parameters of WR\,138a, we use the Potsdam Wolf-Rayet (PoWR) models
for expanding stellar atmospheres. These models account for complex
model atoms including iron-line blanketing in non-LTE (for a
detailed description see Hamann \& Gr\"{a}fener 2004). For
abundances of trace elements, we adopt values which are typical for
Galactic WN stars -- N: 0.015, C: 0.0001, Fe: 0.0014 (Hamann \&
Gr\"{a}fener 2004).

The main parameters of a WR-type atmospheres are the stellar
temperature, $T_\ast$, and the so-called transformed radius, $R_{\rm
t}$. The stellar temperature $T_*$ denotes the effective temperature
related to the radius $R_\ast$, i.e.\ $L = \sigma T_\ast^4 4 \pi
R_\ast^2$, where $\sigma$ is the Stefan-Boltzmann constant and
$R_\ast$ is by definition at a Rosseland optical depth of 20.
$R_{\rm t}$ is related to the mass-loss rate $\dot{M}$ and defined
by
\begin{eqnarray}
R_{\rm t} = R_*
  \left[\frac{v_\infty}{2500 \, \kms} \left/
  \frac{\sqrt{D} \dot M}
       {10^{-4} \, \msun \, {\rm yr^{-1}}}\right]^{2/3} \right. ~, \nonumber
\end{eqnarray}
where $D$ is the clumping contrast (adopted here to be 4 throughout
the wind as a typical value for WN stars; see Hamann \& Koesterke
1998), and $v_\infty$ is the terminal velocity of the wind.

These basic stellar parameters $T_\ast$ and $R_{\rm t}$ are
determined from fitting the lines in the normalized spectrum (see
Fig.\,\ref{fig:linefit}). The normalization is in fact achieved by
dividing the absolutely calibrated observed spectrum by the
theoretical continuum, which makes the total procedure described in
the following an iterative process.

A first orientation about the proper choice of these parameters can
be obtained by comparing the observed spectrum to the published
grids of WN models (Hamann \& Gr\"afener 2004). For a detailed fit,
we calculate individual models for this star with the adequate
terminal wind velocity, which is $v_\infty = 700\,\kms$ as inferred
from fitting the width of the emission line profiles.

The stellar temperature is adjusted such that the balance between
the lines from He\,{\sc i} versus He\,{\sc ii} is reproduced. At the
same time, $R_t$ is adjusted, which influences the strength of the
emission lines in general. The parameters of the best-fitting model
are compiled in Table\,\ref{tab:model} (where for the completeness
we also give the hydrogen ionising luminosity, $\Phi _{\rm i}$).
Note that according to its location in the $T_\ast - R_{\rm t}$
plane (Hamann \& Gr\"{a}fener 2004), WR\,138a belongs to a spectral
subtype later than WN8 (cf. Section\,\ref{sec:type}).

Fig.\,\ref{fig:linefit} shows that the Balmer lines of hydrogen
dominate their blend with lines of the He\,{\sc ii} Pickering
series. Radial velocities of the Balmer lines confirm this. These
blends are nicely matched by models with a hydrogen mass fraction of
20 per cent, leaving about 80 per cent of the mass for helium which
is a typical composition for WNL stars.

The spectral analysis alone cannot tell the absolute dimensions of
the star. These are related to the distance of the object, and will
be discussed in the following subsection.


\subsection{Luminosity and distance of WR\,138a}
\label{sec:dist}

The lowest plausible luminosity for a massive WNL star is about
$\log L/L_{\odot} = 5.3$ (Hamann, Gr\"{a}fener \& Liermann 2006).
Let us adopt this value for a moment. This choice implies the values
for the stellar radius and mass-loss rate as given in
Table\,\ref{tab:model}. Note that, in fair approximation, WR models
can be scaled to different luminosities, with $R_\ast \propto
L^{1/2}$, $\dot{M} \propto L^{3/4}$, and $d \propto L^{1/2}$
(Schmutz, Hamann \& Wessolowski 1989; Hamann \& Koesterke 1998).

\begin{figure}
\begin{center}
\includegraphics[width=1.0\columnwidth,angle=0]{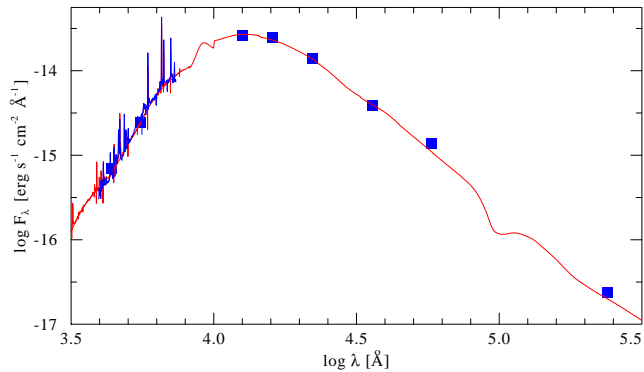}
\end{center}
\caption{Observed flux distribution of WR\,138a (blue/noisy) in absolute units,
including the calibrated spectrum and the photometric
measurements compiled in Table\,\ref{tab:obs}, compared to the emergent flux
of the model continuum (red/smooth line), in the optical also shown with lines.
The model flux has been reddened and scaled to the distance according to
the parameters given in Table\,\ref{tab:model}.
}
\label{fig:SED}
\end{figure}

To estimate the corresponding distance to WR\,138a, we now fit its
observed spectral energy distribution (SED) with the model SED
(Fig.\,\ref{fig:SED}). Two parameters can be adjusted for the fit,
the distance $d$ and the reddening $E_{B-V}$. The reddening law we
adopt from Seaton (1979) in the optical and from Moneti et al.
(2001) in the IR. A perfect fit to the whole SED is achieved with
$E_{B-V} = 2.4\,{\rm mag}$, implying an extinction in the visual of
$A_V = 7.4$\,mag and $d=4.2$ kpc. The strong interstellar absorption
also shows up in the observed spectrum by pronounced DIBs (see
Fig.\,\ref{fig:spec}).

The line-of-sight towards WR\,138a is nearly tangential to the local
(Orion) spiral arm, whose extent in this direction is $\simeq 4-6$
kpc (McCutcheon \& Shuter 1970; Russeil 2003). A distance of about
4\,kpc would imply that this star is located at the far end of the
Orion arm, which is compatible with the high reddening.

Alternatively, one might consider if our object is in fact not a
massive star with ring nebula, but a planetary nebula (PN). PN
central stars with WN-type spectra are very rare; only two of them
are known in our Galaxy [PMR5, Morgan, Parker \& Cohen (2003) and
PB\,8, Todt et al. (2009)] and an eruptive one in the LMC [N\,66;
Hamann et al. (2003)]. With a typical luminosity of a PN central
star, 6000\,$L_\odot$, the photometric distance would become only
0.5\,kpc, placing the object into the foreground of the Orion arm.
In this case, the strong reddening would have to be of circumstellar
origin. Although we cannot strictly rule out this scenario, it
appears very artificial and unlikely.

The distance of $d = 4.2$ kpc, which relies on the adopted stellar
luminosity of $\log L/L_\odot = 5.3$, places WR\,138a at a height of
230\,pc above the Galactic plane. As we will discuss in the next
section, this is just compatible with the inference that WR\,138a is
a runaway star that has been ejected from the Galactic plane. Many
WNL stars have a much larger luminosity than $\log L/L_\odot = 5.3$
(Hamann et al.\ 2006). A higher luminosity, however, would increase
the implied distance ($d \propto L^{1/2}$) and hence lead to an even
larger height above the Galactic plane, while on the other hand the
evolutionary lifetime decreases with higher luminosity. Both effects
together tend to make the runaway scenario impossible for much
larger $d$ and $L$ (see Sect.\,\ref{sec:run}). Hence, we think that
our choice of $\log L/L_\odot = 5.3$ is the most plausible one.

\begin{table}
  \caption{Stellar parameters for WR\,138a}
  \label{tab:model}
  \begin{center}
    \begin{tabular}{lc}
      \hline
      \hline
      Spectral type          &   WN8-9h \\
      $T_\ast$ \, [kK]      &   40     \\
      $\log R_{\rm t} \, [R_\odot ]$ & 0.9 \\
      $v_\infty \, [\kms]$   &   700    \\
      $\log L \, [L_\odot ]$ &   5.3    \\
      $R_\ast \, [R_\odot ]$ &   9.4   \\
      $\log \dot{M} \, [\msun {\rm yr}^{-1} ]$ & -4.7 \\
      $E_{B-V}$ \, [mag]     &   2.4    \\
      $d$ [kpc]              &   4.2   \\
      $\log \Phi _{\rm i} \, [{\rm s}^{-1}]$ & 48.9 \\
      \hline
      \end{tabular}
  \end{center}
\end{table}

\section{Discussion}
\label{sec:dis}

\subsection{WR\,138a as a runaway star}
\label{sec:run}

Our proper motion measurements for WR\,138a
(Section\,\ref{sec:obs-prop}) showed that the star is moving in the
west-southwest direction (i.e. away from the Galactic plane) with a
peculiar (transverse) velocity of $\simeq 50 \, \kms$ (for $d=4.2$
kpc), which is typical of runaway stars. In the Galactic
coordinates, the components of the peculiar velocity are almost
equal to each other: $v_l \simeq -36 \, \kms$, $v_b \simeq 34 \,
\kms$.

Assuming that the progenitor star of WR\,138a was born near the
Galactic plane and ejected from the parent star cluster soon after
the birth, one has that the time it needs to reach the present
height above the Galactic plane, $t_\ast \simeq d\sin b /v_b$,
should be $\leq t_{\rm H}$, where $t_{\rm H}$ is the H-burning
lifetime of the star. Adopting the minimum plausible luminosity for
WNL stars of $\log (L/L_{\odot})=5.3$, one finds the current mass of
WR\,138a of $\simeq 13 \, \msun$, which in turn implies the initial
mass of the star of $\simeq 30 \, \msun$ and $t_{\rm H} \simeq
6.0-7.0$ Myr (e.g. Meynet \& Maeder 2003). Thus, one has that
$t_\ast (\simeq 6.7 $ Myr) $\la t_{\rm H}$, so that the star had
enough time to reach its present location.

Placing WR\,138a at, for example, $d=6$ kpc would result in a
twofold increase of the stellar luminosity and in a increase of the
present and initial mass of the star to, respectively, $\simeq 17 \,
\msun$ and $\simeq 40 \, \msun$. In this case, $t_\ast \simeq 7.0 $
Myr $> t_{\rm H} \simeq 4.5-5.5$ Myr. Further increase of $d$ makes
the discrepancy between $t_\ast$ and $t_{\rm H}$ more severe. The
discrepancy would be even more sever if the star spent several Myr
in the parent cluster before it was ejected (Gvaramadze \& Bomans
2008b; Gvaramadze, Gualandris \& Portegies Zwart 2009). From this
follows that the most likely initial mass of the progenitor star of
WR\,138a was $\simeq 25-30 \, \msun$. One cannot, however, exclude a
possibility that the progenitor star was a blue straggler formed via
merging of two less massive stars in the course of close
binary-binary encounter, resulting in ejection of the merger product
from the parent cluster (e.g. Gvaramadze \& Bomans 2008b). In this
case, the distance to and the mass of WR\,138a could be larger.

The inference that WR\,138a is a runaway star could also be used to
understand the origin of its ring nebula.

\subsection{Origin of the nebula around WR\,138a}
\label{sec:nebula}

A runaway wind-blowing (e.g. WR) star moving through the ISM creates
a highly elongated bubble (Weaver et al. 1977; Brighenti \& D'Ercole
1994), whose shape in the upstream direction is determined by a ram
pressure balance between the stellar wind and the ISM (Baranov,
Krasnobaev \& Kulikovskii 1971). The characteristic scale of the
leading edge of the bubble is $R_{\rm WR} \simeq (\dot{M} _{\rm WR}
v_{\rm WR} /5.6\pi m_{\rm H} n_0 v_{\ast} ^2 )^{1/2}$, where
$\dot{M} _{\rm WR}$ and $v_{\rm WR}$ are the stellar mass-loss rate
and the wind velocity, $n_0$ is the number density of the ambient
medium, $v_{\ast}$ is the stellar peculiar velocity and $m_{\rm H}$
is the mass of a hydrogen atom. For $\dot{M} _{\rm WR} = 10^{-4.7}
\, \msun {\rm yr}^{-1}$ and $v_{\rm WR} = 700 \, \kms$ (see
Table\,\ref{tab:model}), $v_{\ast} = 50 \, \kms$ and $n_0 \simeq 0.1
\, {\rm cm}^{-3}$ (typical of the ISM at $z=230$ pc; Dickey \&
Lockman 1990), one finds $R_{\rm WR} \simeq 20$ pc. The small size
and the nearly circular shape of the nebula around WR\,138a (at
$d=4.2$ kpc, the linear radius of the nebula $r_{\rm neb} \simeq
1.4$ pc) imply that the stellar wind interacts with the dense
ambient medium comoving with the star (cf. Lozinskaya 1992). This
consideration suggests that the immediate precursor of WR\,138a was
a red supergiant (RSG) star (i.e. the initial mass of the progenitor
star was $\leq 40\, \msun$; cf. Section\,\ref{sec:run}) and that the
WR wind still propagates through the region occupied by the dense
material shed during the RSG phase.

According to stellar evolutionary models, a massive star with the
initial mass in the range from $25$ to $40 \, \msun$ evolves through
the sequence O $\rightarrow$ RSG $\rightarrow$ WN (e.g. Meynet \&
Maeder 2003). During the RSG phase, the star loses a considerable
fraction of its initial mass in the form of slow, dense wind. The
subsequent fast WR wind sweeps up the RSG wind and creates a
circumstellar shell. For a spherically symmetric RSG wind, the shell
driven by the WR wind expands with a constant velocity (e.g.
Chevalier \& Imamura 1983) $v_{\rm sh} \simeq (\dot{M} _{\rm WR}
v_{\rm WR} ^2 v_{\rm RSG} /3\dot{M} _{\rm RSG} )^{1/3}$, where
$\dot{M} _{\rm RSG}$ and $v_{\rm RSG}$ are, respectively, the
mass-loss rate and the wind velocity during the RSG phase. Adopting
$\dot{M} _{\rm RSG} =3\times 10^{-5} \msun {\rm yr}^{-1}$ and
$v_{\rm RSG} =10 \, \kms$ (the figures typical of RSGs), one has
$v_{\rm sh} \simeq 100 \, \kms$, which in turn gives us the
dynamical age of the nebula $r_{\rm neb} /v_{\rm sh} \simeq 1.4
\times 10^4$ yr. The latter estimate suggests that WR\,138a only
recently entered the WR phase.

The runaway nature of WR\,138a allows us to propose a natural
explanation of the brightness asymmetry of the nebula and the offset
of the star towards the brightest portion of the nebula. Namely, we
suggest that both are caused by the effect of the ram pressure of
the ISM, which affects the zone occupied by the RSG wind by making
it denser and more compact ahead of the star. This suggestion could
be supported by an estimate of the upstream (minimum) size of the
zone occupied by the RSG wind, $R_{\rm RSG} \simeq (\dot{M} _{\rm
RSG} v_{\rm RSG} /5.6\pi m_{\rm H} n_0 v_{\ast} ^2 )^{1/2}$ ($\simeq
1.6$ pc for the parameters adopted above) $\sim r_{\rm neb}$, which
shows that although the WR wind still propagates through the zone
occupied by the RSG wind, it is already close to the edge of this
zone to fill the density enhancement caused by the stellar motion.
An additional support for our suggestion comes from the proper
motion measurements for WR\,138a, which show that the star is moving
in the direction implied by the above consideration.

Our interpretation of the nebula around WR\,138a as a circumstellar
one (i.e. created via the wind-wind interaction) is consistent with
the observational fact that small-scale (circumstellar) nebulae are
predominantly associated with WNL stars (e.g. Lozinskaya \& Tutukov
1981; Esteban et al. 1993; Barniske, Oskinova \& Hamann 2008), i.e.
with young WR stars, whose winds interact with the material lost
during the preceding evolutionary phases rather than with the
ambient ISM. The scarcity of circumstellar nebulae around WR stars
suggests that they are visible on a time-scale much shorter than the
duration of the WR phase. It is conceivable to associate this
time-scale with the crossing-time of the region occupied by the RSG
wind, i.e. $\simeq r_{\rm RSG} /v_{\rm sh} \la 2-5\times 10^4$ yr,
where $r_{\rm RSG} = v_{\rm RSG} t_{\rm RSG} \simeq 2-5$ pc and
$t_{\rm RSG} \simeq 0.2-0.5$ Myr is the duration of the RSG phase.

\section{Summary}

We have serendipitously discovered a ring nebula around a candidate
WR star, HBHA\,4202-22, in Cygnus using the MIPS $24 \, \mu$m data
from the {\it Spitzer Space Telescope} archive. Our spectroscopic
follow-up confirmed the WR nature of this star and showed that it
belongs to the WN8-9h subtype (we named the star WR\,138a). We
analyzed the spectrum of WR\,138a by using the Potsdam Wolf-Rayet
(PoWR) model atmospheres, obtaining a stellar temperature of 40\,kK.
The stellar wind composition is dominated by helium with 20 per cent
of hydrogen. The stellar spectrum is highly reddened and absorbed
($E_{B-V} = 2.4$\,mag, $A_V = 7.4$\,mag). Adopting a stellar
luminosity of $\log L/L_{\odot} = 5.3$, the star has a mass-loss
rate of $10^{-4.7} \, \msun \, {\rm yr}^{-1}$, and resides in a
distance of 4.2 kpc. We measured the proper motion for WR\,138a and
found that it is a runaway star with a peculiar velocity of $\simeq
50 \, \kms$. The runaway nature of WR\,138a was used to constrain
the initial mass of its progenitor star and to propose an
explanation of the origin of its circular nebula. We found that the
most likely initial mass of the progenitor star was $25-30 \, \msun$
(i.e. the immediate precursor of WR\,138a was a RSG star) and
suggested that the nebula around WR\,138a is a circumstellar one,
created via the interaction between the WR wind and the dense
material shed during the preceding RSG phase.

\section{Acknowledgements}

We are grateful to A.Y.Kniazev for critically reading the manuscript
and to the anonymous referee for useful suggestions. VVG and DJB
acknowledge financial support from the Deutsche
Forschungsgemeinschaft (grants 436 RUS 17/104/06 and BO 1642/14-1)
for research visits of VVG at the Astronomical Institute of the
Ruhr-University Bochum. SF, OS and AFV acknowledge support from the
RFBR grants N\,07-02-00909 and 09-02-00163. AMC acknowledges support
from the RFBR grant N\,08-02-01220 and the State Program of Support
for Leading Scientific Schools of the Russian Federation (grant
NSh-1685.2008.2). This work is based in part on archival data
obtained with the {\it Spitzer Space Telescope}, which is operated
by the Jet Propulsion Laboratory, California Institute of Technology
under a contract with NASA, and has made use of the NASA/IPAC
Infrared Science Archive, which is operated by the Jet Propulsion
Laboratory, California Institute of Technology, under contract with
the National Aeronautics and Space Administration, the SIMBAD
database and the VizieR catalogue access tool, both operated at CDS,
Strasbourg, France.

\end{document}